\PassOptionsToPackage{top=0.75in, bottom=1.09in, left=0.65in, right=0.65in}{geometry}
\documentclass[conference]{IEEEtran}
\IEEEoverridecommandlockouts

\usepackage{cite}
\usepackage{amsmath,amssymb,amsfonts}
\usepackage{graphicx}
\usepackage{textcomp}
\usepackage{xcolor}
\usepackage{url}
\usepackage{breakurl}
\usepackage{geometry}

\usepackage{algorithm}
\usepackage{algpseudocode}
\usepackage{tabularx} 
\usepackage{booktabs} 
\usepackage{listings}
\usepackage[hidelinks,bookmarks=false]{hyperref} 
\hypersetup{draft} 
\usepackage[top=0.75in, bottom=1.09in, left=0.65in, right=0.65in]{geometry}

\algrenewcommand\algorithmicindent{1.0em}
\def\BibTeX{{\rm B\kern-.05em{\sc i\kern-.025em b}\kern-.08em
    T\kern-.1667em\lower.7ex\hbox{E}\kern-.125emX}}
\begin{document}

\title{Beyond Algorithmic Proofs: Towards Implementation-Level Provable Security}

\author{
\IEEEauthorblockN{1
\textsuperscript{st} Jiahui Shang}
\IEEEauthorblockA{
\textit{School of Cyberspace Security} \\
\textit{Communication University of China}\\
Beijing, China \\
hui@cuc.edu.cn
}
\and
\IEEEauthorblockN{2
\textsuperscript{nd} Luning Zhang}
\IEEEauthorblockA{
\textit{School of Cyberspace Security} \\
\textit{Communication University of China}\\
Beijing, China \\
lucyline@cuc.edu.cn
}
\and
\IEEEauthorblockN{3
\textsuperscript{rd} Zhongxiang Zheng*\thanks{* Corresponding author.}}
\IEEEauthorblockA{
\textit{School of Cyberspace Security} \\
\textit{Communication University of China}\\
Beijing, China \\
zhengzx@cuc.edu.cn
}
}

\maketitle
\begin{abstract}
While traditional cryptographic research focuses on algorithm-level provable security, many real-world attacks exploit weaknesses in system implementations, such as memory mismanagement, poor entropy sources, and insecure key lifecycles. Existing approaches address these risks in isolation but lack a unified, verifiable framework for modeling implementation-layer security.
In this work, we propose Implementation-Level Provable Security, a new paradigm that defines security in terms of structurally verifiable resilience against real-world attack surfaces during deployment. To demonstrate its feasibility, we present SEER (Secure and Efficient Encryption-based Erasure via Ransomware), a file destruction system that repurposes and reinforces the encryption core of Babuk ransomware. SEER incorporates key erasure, entropy validation, and execution consistency checks to ensure a well-constrained, auditable attack surface.
Our evaluation shows that SEER achieves strong irrecoverability guarantees while maintaining practical performance. This work demonstrates a shift from abstract theoretical models toward practically verifiable implementation-layer security.
\end{abstract}


\section{Introduction}

Cryptography serves as a cornerstone of information security, with one of its primary goals being the construction of systems resilient to adversarial attacks. Traditional provable security relies on idealized adversarial models and complexity-theoretic reductions, establishing security guarantees under the assumption that breaking the system requires solving well-known hard problems \cite{ProvableSecurityWiki}. However, these models often abstract away critical implementation details—such as randomness generation, memory management, and hardware behavior—and thus fail to account for common practical vulnerabilities, including side-channel leakage, weak entropy sources, and memory corruption. In practice, many real-world breaches stem from such Implementation-Level weaknesses, rendering theoretically secure systems vulnerable once deployed. Consequently, provable security in the theoretical sense is insufficient. There is a pressing need to shift the paradigm toward \textbf{\textit{Implementation-Level Provable Security}}, ensuring that cryptographic systems remain robust not only in abstract models but also in real-world deployments.

One prominent domain where this gap between theory and practice becomes particularly evident is secure file destruction. Ensuring that deleted sensitive data becomes truly irrecoverable—even under forensic analysis or targeted attacks—poses significant challenges. Although various approaches such as overwriting\cite{Liu2013,Yang2022,yingjian}, logical destruction\cite{Diesburg2010}, and cryptographic erasure\cite{Yu2018} have been proposed and standardized, they continue to exhibit serious shortcomings in actual deployment, particularly at the implementation level. These limitations underscore the urgent and representative need for \textbf{\textit{Implementation-Level Provable Security}} in the context of secure file destruction.

We propose SEER (Secure and Efficient Encryption-based Erasure via Ransomware), a system designed to achieve Implementation-Level provable secure file destruction. The core design principle lies in abstracting the empirically validated irrecoverability of ransomware encryption mechanisms and repurposing it into a file destruction scheme with provable Implementation-Level guarantees.

SEER integrates the triple-layer encryption architecture of the Babuk ransomware family, including:
\begin{itemize}
\item Curve25519-based key exchange
\item SHA-256-based key derivation
\item Sosemanuk stream cipher
\end{itemize}

Immediately after use, all keys are securely wiped from memory, ensuring that encrypted data remains unrecoverable.

Unlike conventional approaches that rely solely on theoretical cryptographic strength, SEER explicitly addresses Implementation-Level attack surfaces—such as weak key management, coding flaws, and environmental dependencies. It is designed with practical adversaries in mind, and offers verifiable security guarantees not only in theory but also in real-world deployments.

The primary contributions of this paper are as follows:

\begin{itemize}
\item We formalize the notion of \textbf{\textit{Implementation-Level Provable Security}} and apply it to secure destruction;
\item We design and implement SEER using battle-tested ransomware techniques to ensure data irrecoverability;
\item We propose a dual-layer validation framework integrating formal proof and empirical analysis to verify Implementation-Level robustness.



\end{itemize}

The structure of this paper is as follows: Chapter 2 introduces the concept and background of \textbf{\textit{Implementation-Level Provable Security}}; Chapter 3 presents the design and implementation of the SEER system; Chapter 4 provides a dual-layer security analysis from both theoretical and implementation perspectives; and Chapter 5 concludes the paper and discusses future work.
\addtolength{\topmargin}{0.04in}
\section{Preliminaries}

\subsection{Motivation: Real-World Failures of Cryptographic Implementations}
Modern cryptography offers a strong theoretical foundation for data protection, with algorithms like AES and RSA considered secure under complexity-theoretic assumptions. However, real-world attacks often exploit flaws in implementation rather than the algorithms themselves. Black-box abstractions in theoretical models simplify analysis but overlook critical details such as randomness generation, memory management, and key lifecycle handling\cite{ProvableSecurityWiki}. This gap between theory and implementation is a major source of practical vulnerabilities in cryptographic systems\cite{B3}.


For example, the Heartbleed vulnerability in OpenSSL allowed attackers to read sensitive memory due to missing bounds checks—an implementation bug rather than a flaw in the TLS protocol itself\cite{B6,B7,B8}. Similarly, the 2008 Debian OpenSSL incident, caused by mishandled entropy sources, drastically reduced the randomness of generated keys, making encrypted sessions easily breakable\cite{B10,B12}. Side-channel attacks further demonstrate how physical characteristics like timing, power consumption\cite{B14,B16}, or cache access patterns\cite{B14,B19} can leak secret information, bypassing the assumptions of theoretical models\cite{B21,B25}.

These examples reveal a critical insight: the dominant security risks in real-world cryptographic systems often stem from implementation flaws, not from weaknesses in the theoretical design of cryptographic algorithms. If theoretical security guarantees cannot be reliably preserved in practice, they lose their protective value in real deployments. To improve the robustness of cryptographic systems in operational environments, it is imperative to incorporate a systematic understanding and modeling of Implementation-Level vulnerabilities into the system design process.

\subsection{Definition: Implementation-Level Provable Security}

We introduce the notion of \textbf{\textit{Implementation-Level Provable Security}}, aiming to establish a security reasoning framework that captures critical deployment-stage vulnerabilities, enforces verifiable constraints, and aligns closely with real-world attack surfaces. We define a cryptographic system to possess \textbf{\textit{Implementation-Level Provable Security}} if it satisfies the following two conditions:
\begin{itemize}
\item Its core implementation mechanisms must exhibit a describable and independently verifiable security structure, capable of systematically defending against known or high-probability attack vectors observed in practice.

\item Its deployed version must provide sufficient behavioral consistency checks to ensure that critical execution paths do not deviate from the intended security baseline, and that any potential deviations can be detected through structured verification mechanisms.
\end{itemize}

This definition emphasizes the concept of structural verifiability at the implementation level, rather than aiming for exhaustive formal proofs. Within this framework, we focus on the following core technical dimensions:

\begin{itemize}
\item Memory boundary and information lifecycle control (MemLife), such as secure handling of key generation, usage, and destruction \cite{B8,B9,B35};

\item Observability constraints on execution (ObsGuard), designed to mitigate side-channel leakages (e.g., timing differences and cache access patterns) \cite{B21,B26};

\item Randomness assurance mechanisms (RandCheck), including entropy source quality and compliance of generation workflows;

\item Version consistency validation (VeriHash), such as integrity checking based on code hashes or build fingerprints.

\end{itemize}

Each of these elements has been empirically validated as a core risk factor in major implementation-layer vulnerabilities, making them suitable for systematic modeling and engineering intervention.

SEER embodies this principle by reusing the Babuk ransomware’s encryption and key lifecycle logic. Once data is encrypted and the key securely destroyed, recovery becomes practically impossible—a property SEER repurposes to guarantee data irrecoverability.

In summary, \textbf{\textit{Implementation-Level Provable Security}} offers a new approach to constructing deployment-stage security guarantees that balance theoretical rigor with engineering practicality. It is particularly well-suited for applications where irrecoverability is non-negotiable, such as secure data erasure.

\subsection{Limitations of Existing File Erasure Approaches}

Despite extensive research and established standards, mainstream data destruction methods still face serious Implementation-Level security challenges in real-world deployments.

Data overwriting, one of the oldest and most widely adopted methods, aims to erase residual data by repeatedly writing random bits. For example, the DoD 5220.22-M standard recommends at least three passes\cite{Liu2013,Yu2018}. However, in SSDs, this method is unreliable due to the opaque behavior of the flash translation layer (FTL) and charge interference in 3D NAND flash, leaving 4\%–75\% of data potentially intact\cite{yingjian}. It is also inefficient—a single overwrite of 500GB can take 1.2 hours, and full compliance may exceed 3.6 hours\cite{Arch2024}.

Logical destruction simply removes file system references without erasing actual data\cite{Yang2022,ZenkSecurity}. While fast, this method leaves data fully recoverable using basic forensic tools, offering virtually no real security.

Encryption-based destruction has gained attention for its promise of irrecoverability by encrypting data and deleting the key\cite{Peterson2005a,Peterson2005b}. However, most proposals remain theoretical, relying on cryptographic proofs without real-world validation\cite{Diesburg2010,BigDataNews}. In practice, vulnerabilities in key management, cache remnants, or lack of side-channel protections can compromise the system—even when using strong algorithms like AES\cite{Yu2018}. These risks stem not from flaws in AES itself, but from missing safeguards in the implementation.

In contrast, SEER leverages ransomware-style encryption—already proven effective in hostile environments—to bypass common implementation pitfalls and offer a practical, empirically grounded framework for provably irreversible file destruction.

\section{Protocol Design and Practice of SEER: Secure Erasure via Ransomware Repurposing}

\subsection{Protocol Architecture Based on Ransomware Encryption Mechanisms}

\begin{figure*}[t]  
  \centering
  \includegraphics[width=0.95\textwidth, keepaspectratio]{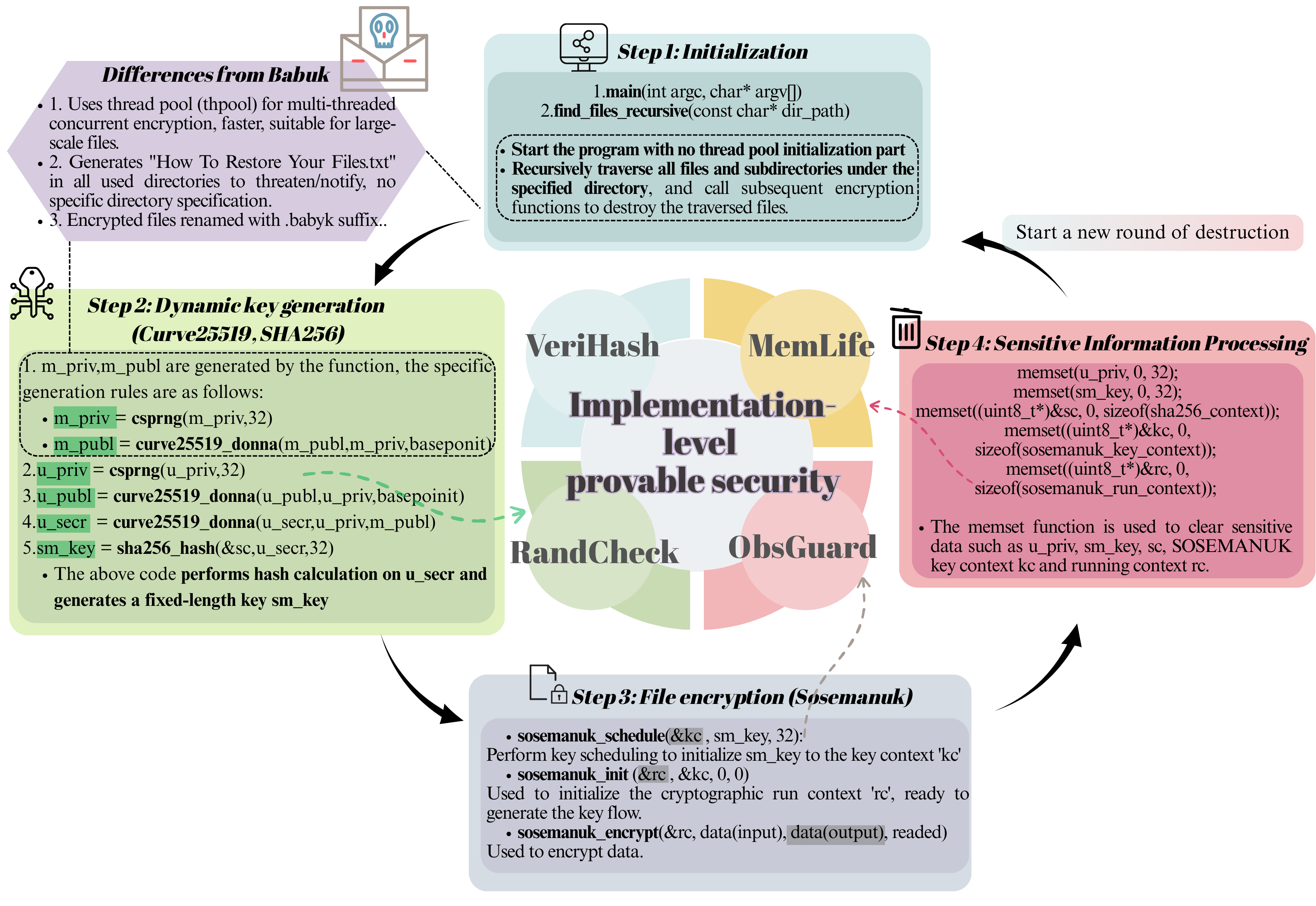}
  \caption{SEER System Workflow and Implementation-Level Security Modeling (Abbreviations: MemLife = Memory \& Lifecycle, ObsGuard = Observability Guard, RandCheck = Randomness Check, VeriHash = Version Hash Verification).}

  \label{fig:SEER System Workflow}
\end{figure*}

The SEER system repurposes the core encryption logic of the Babuk ransomware to construct a three-phase secure destruction protocol: \textbf{ephemeral key generation – file encryption – key destruction}. The key innovation lies in reengineering Babuk's adversarial encryption behavior into a controllable, security-oriented data erasure tool, underpinned by cryptographic protocols and a distributed verification framework (see Figure.~\ref{fig:SEER System Workflow}). Unlike conventional encryption schemes derived from theoretical security assumptions, Babuk's encryption primitives have been empirically validated under real-world adversarial pressure, demonstrating reliable irreversible locking in practice.

SEER inherits several essential components from Babuk's triple-encryption design: Curve25519-based key exchange, SHA-256-based key derivation, and Sosemanuk stream cipher encryption.

In the system's key exchange phase, both ephemeral private keys (\texttt{u\_priv} and \texttt{m\_priv}) are generated using a cryptographically secure pseudorandom number generator (CSPRNG), while the corresponding public keys (\texttt{u\_publ} and \texttt{m\_publ}) are derived using the \texttt{curve25519\_donna()} function. The resulting shared secret \texttt{sm\_key} is computed from \texttt{m\_publ} and \texttt{u\_priv}, then further processed with SHA-256 to produce a session key. This key is used to encrypt the file content via the Sosemanuk stream cipher. Crucially, upon completion, the system invokes the \texttt{memset} function to securely overwrite memory regions holding \texttt{u\_priv} and \texttt{sm\_key}, ensuring that no residual key material remains in memory.

At the encryption layer, SEER employs the Sosemanuk cipher, which integrates the structural efficiency of SNOW 2.0 with non-linear transformation characteristics typical of stream ciphers. This design achieves high-throughput encryption suitable for large-scale file destruction tasks. During actual operation, the ciphertext is generated by encrypting file data with the SHA-256–processed session key, and the corresponding public key is attached to the output to enable traceability. Furthermore, SEER utilizes the \texttt{fseek} function to rapidly overwrite the original file, minimizing the risk of residual data leakage due to buffering or disk-level caching artifacts.

\subsection{Engineering Implementation and Validation}

In building the file destruction system, we modified Babuk ransomware source code: via static analysis, we isolated its core encryption module, removed malicious propagation components, and retained only its formally verified encryption algorithm as the software's core logic. For key management, we designed a dual-entropy random key generation scheme. Cascading Hardware (HRNG) and Software (PRNG) Random Number Generators enabled true random generation of key pairs (\texttt{m\_priv} and \texttt{m\_publ}) and (\texttt{u\_priv} and \texttt{u\_publ}) during encryption, eliminating hard-coded public key vulnerabilities. Core functions are in the Appendix.

The initial full-scale evaluation was conducted in a Linux environment hosted on the ESXi virtualization platform, which offers high stability, while Linux provides a flexible and secure foundation for testing. The test dataset included various file types, such as text, images, and binary executables. Experimental results demonstrate the system's effectiveness and reliability in handling different types of data, as detailed below:

\begin{itemize}
    \item \textbf{Functionality and Security:} The system supports cryptographic destruction for multiple file types. After processing, files become unreadable and cannot be recovered without a valid key. This irreversible outcome verifies the correctness and security of the destruction mechanism.

    \item \textbf{Efficiency:} While destruction efficiency varies by file type, overall performance remains within acceptable bounds. Executable files are processed most efficiently due to their continuous binary structure, which aligns well with block-based encryption. Image files follow, while text files are relatively slower. Despite these differences, processing times across all types meet practical performance requirements.

    \item \textbf{Overall Advantage:} Compared to traditional methods, the system preserves comparable performance while providing enhanced Implementation-Level security. These properties make it a promising solution for irreversible data destruction, balancing security and efficiency, and offering valuable insights for future research and application in secure file disposal.
\end{itemize}

\begin{table}[htbp]
    \caption{Efficiency of the Provable Secure File Destruction System}
    \scriptsize
    \renewcommand{\arraystretch}{1.05}
    \centering
    \begin{tabular}{>{\centering\arraybackslash}p{1.3cm} 
                    >{\centering\arraybackslash}p{1.3cm} 
                    >{\centering\arraybackslash}p{1.3cm} 
                    >{\centering\arraybackslash}p{1.3cm} 
                    >{\centering\arraybackslash}p{1.3cm}}
        \toprule
        \textbf{File Type (1KB)} & \textbf{Erasable} & \textbf{100 Files(s)} & \textbf{1,000 Files(s)} & \textbf{10,000 Files(s)} \\
        \midrule
        Text (.txt)       & \checkmark & 0.187 & 1.450 & 20.522 \\
        Image (.jpg)      & \checkmark & 0.240 & 1.839 & 18.543 \\
        Executable (.exe) & \checkmark & 0.158 & 1.791 & 14.930 \\
        \bottomrule
    \end{tabular}
    
    \label{tab:erasure_efficiency}
\end{table}

It is worth emphasizing that SEER’s primary contribution does not lie in merely improving existing erasure technologies, but rather in proposing a new paradigm of \emph{Implementation-Level security}: repurposing battle-tested encryption components and shifting the focus of security assurance from theoretical completeness to empirical robustness.

The core insight is that real-world adversarial pressure reveals system fragilities and cultivates resilience in ways that abstract theoretical models often overlook. Since its source code was leaked in 2021, Babuk has been used in dozens of real-world attacks, and its encryption has remained unbroken—testament to its practical strength (see Section~4). SEER leverages this ``malicious origin'' to bootstrap a provably robust destruction protocol. This perspective not only offers a fresh angle for secure erasure design, but also presents a generalizable methodology for bridging theoretical rigor and real-world reliability across broader security applications.

\section{Dual-Layer Security Proof}

In cryptography, proving an algorithm’s security often uses \textit{security reductions}—formally linking a scheme’s security to the hardness of well-studied computational problems. Demonstrating these problems are intractable implies the scheme is unbreakable, reducing cryptographic security to assumptions like integer factorization or discrete logarithms.

This section provides a dual-layer security proof for our file destruction system: theoretically, standard reductions show negligible adversary success under hard problem assumptions; implementationally, model-based analysis links formal properties to real-world attack resilience. This framework combines rigor and practicality, overcoming pure theory’s limits for comprehensive validation.

\subsection{Theoretical Security Proof}

An analysis of the source code of the cryptographic pritives employed in our system reveals that its security rests on the following assumptions:
\begin{itemize}
    \item The hardness of the elliptic curve discrete logarithm problem (ECDLP) as instantiated in the Curve25519 algorithm.
    \item The collision resistance of the SHA-256 hash function.
    \item The pseudorandomness of keystreams generated by the Sosemanuk stream cipher.
\end{itemize}

All three assumptions have been widely studied and supported by extensive cryptanalysis.

For Curve25519, it has been shown that known attacks—such as Pollard's rho algorithm and the kangaroo method—incur a computational cost exponential in the curve size. These methods require a large number of elliptic curve point additions, and the success probability of solving ECDLP on Curve25519 is estimated at most \( 2^{-90} \). Moreover, small subgroup attacks are infeasible due to the prime order of the base point, and batch discrete log attacks offer limited advantage\cite{bernstein2006curve25519}.

SHA-256 is a widely deployed hash function known for its strong one-wayness and collision resistance. Its iterative structure and avalanche property significantly increase the difficulty of finding collisions. While theoretical vulnerabilities exist, practical attacks such as differential or birthday attacks have not succeeded against SHA-256. To date, no efficient method for finding SHA-256 collisions has been discovered\cite{Amanda2023}.

Sosemanuk was specifically designed to generate computationally indistinguishable keystreams and is considered secure against known time-memory-data tradeoff (TMDTP) attacks, including Hellman-type attacks. Breaking Sosemanuk requires attacking the internal Serpent24 cipher, with a complexity estimated at \( 2^{128} \) operations. The best known distinguishers achieve a complexity of \( 2^{256} \), which meets modern security expectations\cite{berbain2008sosemanuk}.

\begin{figure}[htbp]
    \centering 
    \includegraphics[width=0.8\linewidth]{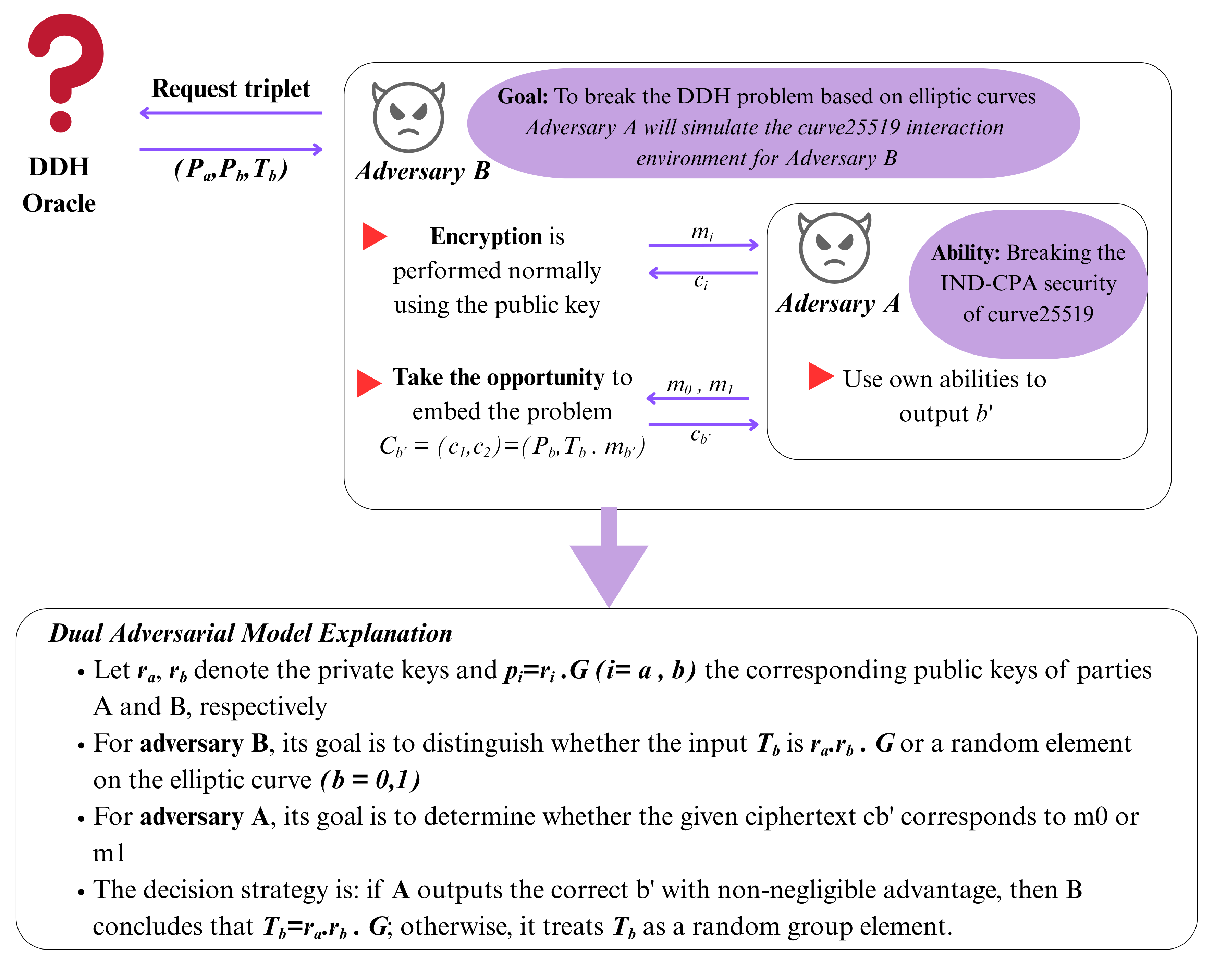} 
    \caption{Double adversary model of Curve25519 algorithm}
    \label{fig:double adversary of Curve25519}
\end{figure}

To illustrate the reduction-based proof approach more concretely, we provide a formal proof for the Curve25519-based key exchange. As shown in the Figure \ref{fig:double adversary of Curve25519}, the target is to establish \textit{semantic security}, i.e., indistinguishability under chosen-plaintext attack (IND-CPA). Consider an adversary \( \mathcal{A} \) attempting to distinguish ciphertexts. A simulator \( \mathcal{B} \) embeds a Decisional Diffie-Hellman (DDH) instance into the challenge and interacts with \( \mathcal{A} \), using its responses to break the DDH assumption.\cite{Max1z2023}

    \textbf{Adversary \( \mathcal{A} \) (System Attacker)}
        \begin{itemize}
            \item \textbf{Attack Objective}: Crack the IND - CPA security of curve25519 and figure out the plaintext corresponding to a given ciphertext.
            \item \textbf{Capability Scope}:
                \begin{itemize}
                    \item Can use its own capabilities to output the judgment result $b'$.
                    \item Is able to interact with adversary \( \mathcal{B} \) and receive the ciphertext $c_{b'}$ sent by adversary \( \mathcal{B} \).
                \end{itemize}
         \end{itemize}

    \textbf{Adversary \( \mathcal{B} \) (Underlying Cryptographic Attacker)}
        \begin{itemize}
            \item \textbf{Attack Objective}: Crack the DDH problem based on the elliptic curve and tell whether the input $T_b$ is $r_a r_b G$ or a random element on the elliptic curve ($b = 0, 1$).
            \item \textbf{Capability Scope}:
                \begin{itemize}
                    \item Allows adversary \( \mathcal{A} \) to simulate the curve25519 interaction environment.
                    \item Can use the public key for normal encryption operations.
                    \item Can embed the problem $C_{b'}=(c_1, c_2)=(P_b, T_b, m_{b'})$ in the encryption process.
                    \item Can request the triple $(P_a, P_b, T_b)$ from the DDH Oracle.
                \end{itemize}
        \end{itemize}

Hence, the difficulty of compromising the key exchange is reduced to the hardness of the DDH problem on the elliptic curve underlying Curve25519, which is a standard approach in provable security.

However, the aforementioned arguments address only the idealized model of the algorithms and do not account for practical implementation issues. Despite the theoretical security of Curve25519, SHA-256, and Sosemanuk, the actual deployment of these algorithms may introduce vulnerabilities that significantly diminish system security. This echoes long-standing concerns in public-key cryptanalysis regarding the gap between theory and practice.

Implementation flaws—coding bugs, parameter truncation, side-channel vulnerabilities—are frequent attack vectors, especially in constrained environments like smart cards. For example, shortening Sosemanuk’s key length for efficiency could reduce attack complexity from \(2^{128}\) to a tractable level, nullifying its design.  

Similarly, Curve25519’s theoretical security is undermined by weak key generation (e.g., poor CSPRNG entropy causing predictable private keys) or side-channel leaks. Timing variations in elliptic curve computations, for instance, may expose private key bits, violating theoretical guarantees.  

These cases confirm a core principle: cryptographic security depends on both underlying problem hardness and implementation rigor. Neglecting implementation details leaves systems vulnerable in practice, despite sound theoretical foundations.

\subsection{Implementation-Level Provable Security}

\begin{figure}[htbp]
    \centering 
    \includegraphics[width=0.8\linewidth]{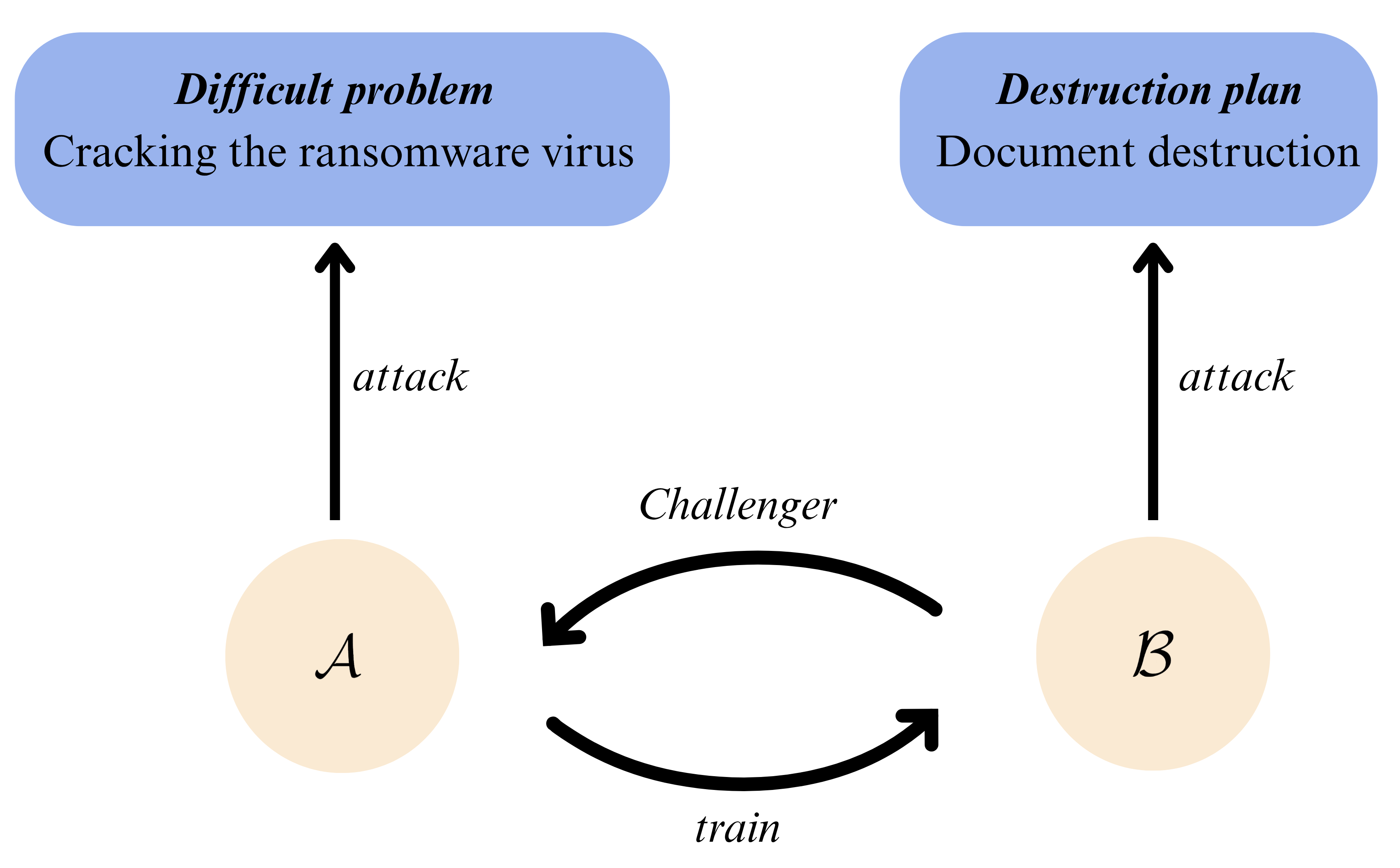} 
    \caption{Dual-Adversary Model used in SEER's Implementation-Level security analysis}
    \label{fig:dual_adversary_model}
\end{figure}

To address the limitations of purely theoretical proofs, SEER extends its security guarantees into the implementation layer by constructing a two-pronged verification framework. The first component introduces a dual-adversary model, illustrated in Figure~\ref{fig:dual_adversary_model}, which leverages the real-world exposure and analysis of the Babuk ransomware to formalize attacker capabilities in practical environments. This model builds upon traditional dual-adversary modeling approaches\cite{Yang2017} while deeply incorporating the empirically validated traits of Babuk—whose source code was publicly leaked in 2021. This disclosure empowers researchers with capabilities effectively equivalent to those of adversary \( \mathcal{B} \) in our model, enabling them to simulate advanced attack scenarios such as chosen-ciphertext attacks. As a result, the adversary’s capabilities are not idealized abstractions, but rather faithful representations of real-world threats grounded in tangible, observable behaviors.

The second component integrates implementation - level formal modeling for security - critical operation validation (memory, randomness, time - constant execution). These approaches bridge theory - practice gaps through adversarial proofs and empirical traceability.

\subsubsection{Dual-Adversary Model}

\textbf{Adversary \( \mathcal{A} \) (System Attacker)}
\begin{itemize}
    \item \textbf{Attack Objective}: Recover the plaintext content of the file destruction system.
    \item \textbf{Capability Scope}:
        \begin{itemize}
            \item Have full control over the operating environment of the file destruction system (white-box access).
            \item Be able to intercept an arbitrary number of plaintext-ciphertext pairs \((m_i, c_i)\).
            \item Be able to tamper with the temporary parameters in the encryption process (such as the nonce value).
        \end{itemize}
\end{itemize}

\textbf{Adversary \( \mathcal{B} \) (Underlying Cryptographic Attacker)}
\begin{itemize}
    \item \textbf{Attack Objective}: Breach the core encryption module of the Babuk ransomware.
    \item \textbf{Capability Scope}:
        \begin{itemize}
            \item Be able to obtain the complete engineering implementation of the virus sample (including the code of Curve25519, SHA - 256, and Sosemanuk).
            \item Be able to inject a custom public key for exchange.
            \item Be able to observe the physical side channels of the encryption process (such as power consumption, electromagnetic radiation).
        \end{itemize}
\end{itemize}

Our core proposition is that if Adversary \( \mathcal{B} \) cannot breach the Babuk encryption module (security assumption), then Adversary \( \mathcal{A} \) cannot breach the file destruction system (the conclusion to be proven). The formal expression is:

\[
\begin{aligned}
&\forall \text{PPT Adversary } \mathcal{A}, \exists \text{PPT Adversary } \mathcal{B}, \\
&\quad \text{Adv}_\mathcal{A} \leq \text{Adv}_\mathcal{B} + \text{negl}(\lambda)
\end{aligned}
\]

(where \(\text{Adv}_\mathcal{B}\) represents the advantage of Adversary \( \mathcal{A} \) in breaching the core encryption module of the ransomware, \(\text{Adv}_\mathcal{A}\) represents the advantage of Adversary \( \mathcal{A} \) in breaching the file destruction system, and \(\text{negl}(\lambda)\) is a negligible function.)

Based on the assumption that files encrypted by Babuk ransomware are irrecoverable without the key, we establish a security reduction: breaking our file destruction system is no easier than breaking Babuk’s encryption. Thus, any successful file recovery implies a feasible attack on the ransomware itself.

\subsubsection{Implementation-Level Security Modeling and Verification}

To verify SEER at the implementation level, we leverage Babuk’s proven resistance to real-world recovery as a foundation for formal analysis. By structurally aligning with Babuk’s encryption modules and validating critical components, SEER achieves a provable and practically robust security baseline.

\textbf{MemLife}. Babuk enforces strict key lifecycle management. Upon completing cryptographic operations, sensitive memory regions are actively zeroized using \texttt{memset} (e.g., \texttt{memset(u\_priv, 0, 32)}), preventing residual data leakage. Static analysis tools can be employed to formally model memory safety, ensuring the secure erasure of cryptographic material.

\textbf{RandCheck}. High-quality randomness is sourced via a \texttt{csprng} that reads directly from \texttt{/dev/urandom}, avoiding weaknesses associated with userspace PRNGs. Combined with Curve25519-compliant constraints (e.g., \texttt{u\_priv[0] \&= 248}), this ensures both entropy quality and key validity. Static and symbolic analysis techniques can be applied to model entropy strength and verify the unpredictability of key generation.

\textbf{ObsGuard}. Babuk processes input using fixed-size data blocks (with \texttt{CONST\_BLOCK} defined as 10MB), eliminating execution-time variance due to input size. This design enables formal reasoning about constant-time behavior and provides a practical foundation for mitigating timing side-channel attacks.

\textbf{VeriHash}. To ensure integrity after integration and adaptation, we perform hash-based verification on code segments against the original Babuk source. This confirms byte-level preservation of core cryptographic logic (including Curve25519, SHA-256, and Sosemanuk), and verifies that no inadvertent modifications were introduced. This mechanism is further extended to compiled binaries, ensuring build-time consistency and enabling precise inheritance and traceability of critical security routines.

Looking forward, SEER can evolve this analysis process into a machine-verifiable proof chain—transitioning from empirically grounded inheritance to fully verified implementations under formal modeling. This fusion of real-world adversarial resilience with cryptographic rigor provides a paradigm for extracting and transforming defensive mechanisms from hostile infrastructure into secure and verifiable foundations.

\subsection{Case-Based Evidence}

To validate the Implementation-Level security of SEER’s encryption, we conduct empirical analysis of Babuk-related ransomware incidents and public decryption efforts.

\subsubsection{Real-World Incident Analysis}

Since Babuk’s source release in 2021~\cite{Lawrence2021}, its core encryption remains unbroken and underpins ransomware families like Play, Lockr, and RaGroup. SentinelOne’s 2023 report~\cite{Alex2023} highlights at least nine active strains still reusing Babuk’s encryption, indicating its robustness.

Table~\ref{tab:babuk_attacks} summarizes major Babuk-related attacks from 2021 to 2024. Notably:

\textbf{Washington D.C. Police (April 2021):} Over 250GB of sensitive data was encrypted and leaked, with no successful decryption despite federal intervention, demonstrating Babuk’s irreversible encryption.

\textbf{Houston Rockets (April 2021):} Around 500GB of confidential documents were encrypted; professional defenses failed to recover data, and ransom was paid, underscoring practical inaccessibility of unauthorized decryption.

\begin{table}[htbp]
    \caption{Babuk-Based Ransomware: Crypto Case Implications\label{tab:babuk_attacks}}
    \scriptsize
    \renewcommand{\arraystretch}{1.05}
    \centering
    \begin{tabular}{>{\centering\arraybackslash}p{1.5cm} 
                    >{\centering\arraybackslash}p{0.6cm} 
                    >{\centering\arraybackslash}p{2.8cm} 
                    >{\centering\arraybackslash}p{2.3cm}}
        \toprule
        \multicolumn{1}{c}{\textbf{Target}} & 
        \multicolumn{1}{c}{\textbf{Date}} & 
        \multicolumn{1}{c}{\textbf{Summary}} & 
        \multicolumn{1}{c}{\textbf{Implication}} \\
        \midrule
        DC Police\cite{Lawrence2021-DC} & 04/2021 & 250GB locked; \$4M ransom; partial leak & Govt unable to decrypt; shows strong irreversibility \\
        Rockets\cite{Urian2021} & 04/2021 & 500GB encrypted; ransom paid & No recovery without ransom; fallback failed \\
        Serco\cite{WHITEPAPER} & 02/2021 & 1TB breach; ops disrupted & Long-term stealth and infiltration \\
        Ra Group\cite{Alex2023} & 04/2023 & Babuk-based; 3 firms hit & Core encryption reused; still secure \\
        Play Variant\cite{Alessandro2023} & 06/2023 & Double extortion; leak threats & Encryption remains critical enabler \\
        \bottomrule
    \end{tabular}
    
\end{table}

These cases collectively validate the real-world intractability of Babuk’s encryption. Even under well-resourced defense contexts, the mechanism maintained its non-recoverability, thus providing SEER with a credible Implementation-Level security foundation.

\subsubsection{Public Decryption Attempts}

The publication of Babuk's codebase prompted intensive analysis by security researchers. In 2021, Avast released a decryption tool applicable only to a narrow subset of cases involving leaked keys, exploiting weaknesses in key management rather than cryptographic primitives\cite{Sergiu2021}. Similarly, a 2024 tool developed jointly by Avast and Cisco targeting the “Tortilla” variant depends on pre-leaked keys or embedded decryptors, amounting to key duplication rather than a fundamental cryptanalytic breakthrough\cite{landiannews2024}.

Crucially, none of these tools target the core cryptographic primitives---Curve25519, SHA-256, or Sosemanuk---nor have any succeeded in circumventing their correct implementation. Even under white-box conditions with full source visibility, no effective attack against the original scheme has been demonstrated. As of this writing, \emph{no public tool exists to decrypt data encrypted by the original Babuk mechanism}, affirming its robust Implementation-Level security posture.

\subsubsection{Summary and Implications}

Taken together, these findings provide strong empirical backing for the provable security claims of SEER at the implementation level.

From an adversarial standpoint, Babuk and its derivatives have launched over 42 documented attacks against diverse targets over multiple years. Despite source code leakage, its core encryption remained uncompromised, affirming its resilience in adversarial environments.

In the context of publicly disclosed cryptanalysis, although attack strategies against Babuk have continued to evolve within the industry, its core encryption mechanism remains robust, with critical procedures adhering to established security standards. Numerous decryption attempts have failed to compromise the fundamental core logic, and existing tools rely solely on a subset of keys leaked during the 2021 source code disclosure or target specific variants (which do not satisfy the VeriHash condition as defined in our model). This underscores the scheme’s controllable and reliable security posture in deployed environments.

SEER harnesses the irreversible properties of Babuk’s encryption---proven in practice---as a constructive security asset. Without needing full control of the implementation internals, SEER builds on the fact that once keys are erased, encrypted data becomes permanently unrecoverable.

Therefore, the assumption that \textit{Adversary \( \mathcal{B} \)} cannot break Babuk’s encryption holds in practice, which in turn implies that \textit{Adversary \( \mathcal{A} \)}---tasked with recovering securely deleted files within bounded time---faces equivalent difficulty. Thus, the provable security of SEER’s file destruction mechanism is upheld at the implementation level.

\section{Conclusion}

This paper addresses the core challenge of bridging theoretical assurances and practical deployment in Implementation-Level provable security. Traditional abstract security models fail to capture vulnerabilities from implementation details such as side-channel leaks and memory management flaws. We introduce \textbf{\textit{Implementation-Level Provable Security}}, defining it as requiring attack resistance verified in real environments, verifiable and structurally resilient critical components, and behavioral consistency checks ensuring key execution paths meet security baselines. The overall system security must reduce to a known, empirically robust primitive, closing the theory-practice gap.

As a proof of concept, we present SEER, a secure file destruction system based on these principles. Experiments validate its practical effectiveness and efficiency.

SEER’s security relies on a dual-layer approach:

\begin{itemize}
    \item Theoretical security founded on cryptographic hardness assumptions of primitives like Curve25519, SHA-256, and Sosemanuk.
    \item Implementation-Level security constructed under a dual-adversary model, reducing its resilience to the real-world unbreakability of Babuk ransomware’s core encryption.
\end{itemize}

SEER's solution realizes Implementation-Level modeling(including mechanisms such as secure key erasure and file overwriting) and reduces its unrecoverability to the unbreakability of the Babuk ransomware to validate its effectiveness, thereby effectively bridging the gap between theoretical security and real-world deployment security.

This demonstrates that the future of secure data erasure lies in such a comprehensive, multi-layered methodology—one that not only relies on the theoretical robustness of cryptographic primitives but also rigorously enforces Implementation-Level safeguards. This advancement signifies an evolution in security paradigms: from abstract theoretical guarantees to concrete, verifiable security in deployed systems, thereby establishing a new pathway for engineering cryptographic security.

\appendix


\renewcommand{\thesection}{A\arabic{section}}
\setcounter{section}{0}
\makeatletter
\renewcommand{\section}{\@startsection{section}{1}%
  \z@{\linespacing\@plus\linespacing}{0.5\linespacing}%
  {\normalfont\large\bfseries}}
\makeatother

\begin{center}
\textbf{A1. Random Number Generation}
\end{center}

\begin{minipage}{0.95\linewidth} \begin{algorithm}[H]
\caption{Cryptographically Secure Random Number Generation}
\label{alg:csprng}
\begin{algorithmic}[1]
\Procedure{CSPRNG}{$\text{buffer}, \text{count}$}
    \State Open \texttt{/dev/urandom} as file $fp$
    \If{$fp \neq$ NULL}
        \State Read $count$ bytes from $fp$ into $\text{buffer}$
        \State Close $fp$
    \EndIf
\EndProcedure
\end{algorithmic}
\end{algorithm} \end{minipage}

This function securely generates high-entropy random bytes by reading from the Linux system’s \texttt{/dev/urandom}. The output is written into \texttt{buffer} and is primarily used for generating cryptographic secrets such as private keys. This mechanism ensures strong unpredictability in key materials.

\begin{center}
\textbf{A2. Recursive File Traversal}
\end{center}

\begin{minipage}{0.95\linewidth} \begin{algorithm}[H]
\caption{Recursive Directory Traversal for File Encryption}
\label{alg:dir_traversal}
\begin{algorithmic}[1]
\Procedure{FindFilesRecursive}{$\text{dir\_path}$}
    \State Open directory at $\text{dir\_path}$ as $dir$
    \If{$dir = \text{NULL}$}
        \State \textbf{return}
    \EndIf
    \ForAll{$\text{entry} \in \text{readdir}(dir)$}
        \If{$\text{entry} = "." \lor \text{entry} = ".."$}
            \State \textbf{continue}
        \EndIf
        \State $full\_path \gets \text{dir\_path} \Vert "/" \Vert \text{entry}$
        \If{entry is directory}
            \State \Call{FindFilesRecursive}{$full\_path$}
        \ElsIf{entry is regular file and not ends with ``.encrypted''}
            \State \Call{EncryptFile}{$full\_path$}
        \EndIf
    \EndFor
    \State Close $dir$
\EndProcedure
\end{algorithmic}
\end{algorithm} \end{minipage}

This utility recursively traverses the specified directory and its subdirectories. For each regular file that does not have a \texttt{.encrypted} suffix, the function invokes \texttt{encrypt\_file} to encrypt its contents. This enables automated bulk processing over an entire file system subtree.

\begin{center}
\textbf{A3. Per-File Encryption Workflow}
\end{center}

\begin{minipage}{0.95\linewidth} 
\begin{algorithm}[H]
\caption{Per-File Authenticated Encryption}
\label{alg:encrypt_file}
\small
\begin{algorithmic}[1]  
\Procedure{EncryptFile}{$\text{path}$}
    \State Generate ephemeral key pair $(u_{priv}, u_{publ})$
    \State Retrieve fixed public key $m_{publ}$
    \State $u_{secr} \gets \text{Curve25519}(u_{priv}, m_{publ})$
    \State $sm_{key} \gets \text{SHA-256}(u_{secr})$
    \State Initialize SOSEMANUK cipher with $sm_{key}$
    \ForAll{block in file at \texttt{path}}
        \State Encrypt block in-place using SOSEMANUK
    \EndFor
    \State Append $u_{publ}$ to file
    \State Rename file with ``.encrypted'' suffix
    \State Securely erase $u_{priv}$, $sm_{key}$, and cipher context
\EndProcedure
\end{algorithmic}
\end{algorithm} 
\end{minipage}
This function performs authenticated encryption of an input file with the following steps:

\begin{enumerate}
    \item \textbf{Key Generation and Exchange}: Two ephemeral Curve25519 key pairs are generated: (\texttt{m\_priv} and \texttt{m\_publ}) and (\texttt{u\_priv} and \texttt{u\_publ}). A Diffie-Hellman key exchange is performed between (\texttt{u\_priv} and \texttt{m\_publ}), resulting in the shared secret \texttt{u\_secr}.
    
    \item \textbf{Key Derivation}: The shared secret \texttt{u\_secr} is hashed via SHA-256 to derive a 256-bit symmetric key \texttt{sm\_key}.
    
    \item \textbf{Stream Cipher Initialization}: The SOSEMANUK stream cipher is initialized via \texttt{sosemanuk\_schedule} and \texttt{sosemanuk\_init} using \texttt{sm\_key}.
    
    \item \textbf{File Encryption}: The file is read in blocks and encrypted in-place using \texttt{sosemanuk\_encrypt}. The original contents are overwritten with ciphertext.
    
    \item \textbf{Key Embedding}: The session public key \texttt{u\_publ} is appended to the end of the encrypted file to support decryption.
    
    \item \textbf{Finalization}: The file is renamed with a \texttt{.encrypted} suffix. All sensitive variables (e.g., private keys, symmetric key, and cipher context) are securely erased from memory using \texttt{memset}.
\end{enumerate}

This process guarantees confidentiality of the file contents and ensures forward secrecy by using per-file ephemeral keys.

\begin{center}
\textbf{A4. SOSEMANUK Stream Cipher Internals}
\end{center}

\begin{enumerate}
    \item \textbf{sosemanuk\_schedule}
    
    \textit{Purpose}: Expands the 256-bit key into the cipher’s internal key context.

\begin{minipage}{0.95\linewidth} \begin{algorithm}[H]
\caption{SOSEMANUK Key Schedule}
\label{alg:sosemanuk_schedule}
\small
\begin{algorithmic}[1]
\Procedure{sosemanuk\_schedule}{$\text{kc}, \text{key}, \text{keylen}$}
    \For{$i \gets 0$ \textbf{to} $7$}
        \State $\text{kc.k}[i] \gets \text{key}[4i] \ |\ (\text{key}[4i+1] \ll 8)\ |\ (\text{key}[4i+2] \ll 16)\ |\ (\text{key}[4i+3] \ll 24)$
    \EndFor
    \State \Call{serpent\_key\_schedule}{$\text{kc.serpent\_keys}, \text{kc.k}$}
\EndProcedure
\end{algorithmic}
\end{algorithm} 
\end{minipage}

First, it split the 32-byte input key into 8 32-bit integers; then, it perform key expansion using the Serpent algorithm to generate round keys; finally, it initialize the key context kc for SOSEMANUK.
    
    \item \textbf{sosemanuk\_init}
    
    \textit{Purpose}: Initializes the cipher’s internal state using the scheduled keys.
    
    \begin{minipage}{0.95\linewidth} \begin{algorithm}[H]
\caption{SOSEMANUK Initialization}
\label{alg:sosemanuk_init}
\begin{algorithmic}[1]
\Procedure{sosemanuk\_init}{$\text{rc}, \text{kc}, \text{iv}, \text{ivlen}$}
    \State \Call{serpent\_init}{$\text{rc.serpent\_state}, \text{kc.serpent\_keys}$}
    \For{$i \gets 0$ \textbf{to} $9$}
        \State $\text{rc.lfsr}[i] \gets \text{kc.k}[i \bmod 8] \oplus (i \times \text{0x9e3779b9})$
    \EndFor
    \State $\text{rc.idx} \gets 0$
\EndProcedure
\end{algorithmic}
\end{algorithm} 
\end{minipage}

It initializes the Serpent cipher state using round keys, seeds the LFSR (Linear Feedback Shift Register) — a core component of SOSEMANUK’s keystream generator — and prepares the runtime context \texttt{rc} for encryption.
    
    \item \textbf{sosemanuk\_encrypt}
    
    \textit{Purpose}: Encrypts (or decrypts) data using a generated keystream.
    
    \begin{minipage}{0.95\linewidth} 
    \begin{algorithm}[H]
\caption{SOSEMANUK Encryption/Decryption}
\label{alg:sosemanuk_encrypt}
\begin{algorithmic}[1]
\Procedure{sosemanuk\_encrypt}{$\text{rc}, \text{in}, \text{out}, \text{len}$}
    \For{$i \gets 0$ \textbf{to} $\text{len}-1$}
        \If{$\text{rc.idx} == 0$}
            \State \Call{generate\_keystream\_block}{$\text{rc}, \text{rc.keystream\_buf}$}
        \EndIf
        \State $\text{ks\_byte} \gets \text{rc.keystream\_buf}[\text{rc.idx}]$
        \State $\text{rc.idx} \gets \text{rc.idx} + 1$
        \If{$\text{rc.idx} == \text{BLOCK\_SIZE}$}
            \State $\text{rc.idx} \gets 0$
        \EndIf
        \State $\text{out}[i] \gets \text{in}[i] \oplus \text{ks\_byte}$
    \EndFor
\EndProcedure
\end{algorithmic}
\end{algorithm} 
\end{minipage}

It produces keystream bytes on demand, XORs each plaintext byte with a keystream byte to generate ciphertext, and — owing to the symmetric nature of stream ciphers — uses the same function for decryption.

\end{enumerate}

\begin{center}
\textbf{A5. Secure Memory Erasure}
\end{center}

All sensitive data (e.g., private keys, derived symmetric keys, cipher contexts) are explicitly cleared from memory using \texttt{memset}. Example calls include:

\begin{minipage}{0.95\linewidth} \begin{algorithm}[H]
\caption{Secure Memory Erasure}
\label{alg:secure_memset}
\begin{algorithmic}[1]
\Procedure{SecureErase}{}
    \State Overwrite $u_{priv}$ with zeros
    \State Overwrite $sm_{key}$ with zeros
    \State Overwrite SHA256 and SOSEMANUK contexts with zeros
\EndProcedure
\end{algorithmic}
\end{algorithm} 
\end{minipage}

This practice mitigates the risk of residual data being recovered from memory after encryption is complete, aligning with secure coding standards.

\end{document}